\documentclass[aps,prl,showpacs,superscriptaddress,twocolumn]{revtex4}
\usepackage{amsmath}
\usepackage{amsfonts}
\usepackage{graphicx}
\usepackage{psfrag}
\begin{document}

\title{Spin-orbit interaction controlled properties of two-dimensional superlattices: Spintronic crystals}
\author{P\'{e}ter F\"{o}ldi}
\email{foldi@physx.u-szeged.hu}
\affiliation{Department of Theoretical Physics, University of Szeged, Tisza Lajos k\"{o}r%
\'{u}t 84, H-6720 Szeged, Hungary}
\author{Viktor Szaszk\'{o}-Bog\'{a}r}
\affiliation{Department of Theoretical Physics, University of Szeged, Tisza Lajos k\"{o}r%
\'{u}t 84, H-6720 Szeged, Hungary}
\author{F. M. Peeters}
\affiliation{Departement Fysica, Universiteit Antwerpen, Groenenborgerlaan 171, B-2020
Antwerpen, Belgium}

\begin{abstract}
The band structure of two-dimensional artificial superlattices in the presence of (Rashba-type) spin-orbit interaction (SOI) is presented. The position and shape of the energy bands in these spintronic crystals depend on the geometry as well as the strength of the SOI, which can be tuned by external gate voltages. For finite mesoscopic arrays we show that their conductance properties can be understood from these spin-dependent band diagrams.
\end{abstract}

\pacs{85.75.-d, 85.35.Ds, 61.50.Ah}
\maketitle

\section{I.~Introduction}
\vspace{-0.5cm}
Infinite periodic structures -- although in the strict sense they do not exist -- provide several fundamental concepts that determine the physical properties of various systems. E.g., the notion of bands is essential in the theory of solids. The concept of "crystals" has also been used e.g. in the context of photonic bandgap materials \cite{Y87,S87}. In that case artificial periodic structures have been fabricated with a photonic band structure. Here we show how both the energy and spin of electrons can be engineered leading to the concept of spintronic crystals.
The band structure of these crystals is flexible in the sense that instead of being completely determined by the geometry, it is tunable by gate voltages.

We consider two-dimensional superlattices, which can be fabricated from e.g. InAlAs/InGaAs based heterostructures \cite{KNAT02} or HgTe/HgCdTe quantum wells \cite{KTHS06}, where Rashba-type \cite{R60} spin-orbit interaction (SOI) is present. This effect, which is essentially the same as the one which causes the fine structure in atomic spectra,
results in spin-dependent transport phenomena. Experiments have demonstrated that the strength of this type of SOI can be controlled by external gate voltages \cite{NATE97,G00}. Small periodic structures like $5\times 5$\ ring arrays have recently been realized experimentally \cite{BKSN06} and have been described theoretically \cite{ZW07,KFBP08b} as well. Finite chains of quantum rings \cite{MVP05}, ladder \cite{WXE06} and diamond-like elements \cite{A08,SMC09} have also been investigated, as well as artificial crystal-like structures \cite{MT09,BO10} in contexts different from the current one. The spin transformation properties of finite networks suggest various possible spintronic \cite{ALS02} applications as well \cite{KFBP08c,A08,KKF09}.

\begin{figure}[tbh]
\includegraphics[width=8cm]{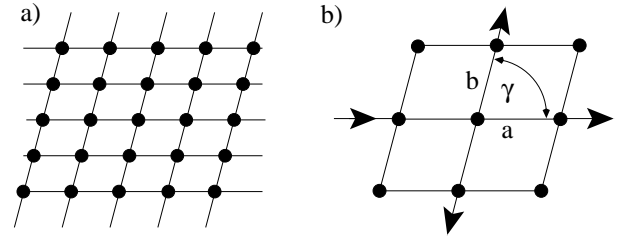}
\caption{Two-dimensional parallelogram lattice (a) and a finite array (b), where the relevant parameters (that may lead to square or rectangular lattices as well) and some possible input and output leads are shown. Electrons  move along the lines connecting the junctions (full circles).}
\label{latticefig}
\end{figure}

Conductance calculations show -- already in relatively small finite lattices -- signatures of a possible underlying band structure, as one can identify SOI-dependent energy regions where the system is completely opaque for the electrons. In the present paper we show that these "non-conducting stripes" \cite{KFBP08b} are directly related to the relevant band gaps. The general two-dimensional (2D) lattice that we will consider is shown in Fig.~\ref{latticefig}. We found that the details of the band structure strongly depends on the geometry of the unit cell, but the overall scaling properties, as well as the remarkable SOI dependence are general.

The characteristic energies in our problem are much smaller than e.g. the usual electronic band gaps in semiconductors. This is essentially due to the differences in the lattice constants, typical "bond lengths" have to be compared to spatial periodicity of the artificial superlattices, which is around 10 nm. Thus the nanometer-scale translational symmetry induced subbands that we describe in the current paper can be considered as a fine structure of the usual electronic bands, which becomes important at low temperatures.

%%%%%%%%%%%%%%%%%%%%%%%%%%%%%%%%%%%%%%%%%%%%%%%%%%%%%%%%%%%%%%%%%%%%%%%%%%%%%%%%%%%%%%%%%%%%%%%%%%%%%%%%%%%%%%%%%%%%%%%%%%%%%%%%%%%%%%%%%%%%%%%%%%%%%%%%%%%%
%%%%%%%%%%%%%%%%%%%%%%%%%%%%%%%%%%%%%%%%%%%%%%%%%%%%%%%%%%%%%%%%%%%%%%%%%%%%%%

\section{II.~Calculation of the spin-dependent band structures}
\vspace{-0.5cm}
The Hamiltonian of a narrow quantum wire in the $x$-$y$ plane with Rashba-type SOI can generally be written \cite{MMK02}  as
\begin{equation}
H = \hbar\Omega\left[ \left( -i\frac{\partial}{\partial s}+\frac {\omega}{2\Omega} \mathbf{n}(\mathbf{\sigma}\times\mathbf{e}_z) \right) ^{2}-\frac{\omega^{2}}{4\Omega^{2}}%
\right],
\label{Ham}
\end{equation}
where the unit vector $\mathbf{n}$ points to the chosen positive direction along the wire, and we introduced the characteristic kinetic energy $\hbar\Omega=\hbar^{2}/2m^{\ast}a^{2}$ (with $a$ being one of the lattice constants, see Fig.~\ref{latticefig}). The strength of the SOI is given by $\omega=\alpha/a,$ where the Rashba parameter $\alpha$ \cite{MMK02} is tunable by gate voltages, and $s$ denotes the (dimensionless) length variable along the wire measured in units of $a.$

Independently from the direction of the wire, the energy levels of $H$ form a continuum, and the spin direction of the eigenspinors depend on the SOI strength. A given energy eigenvalue is fourfold degenerate due to the two possible propagation and spin directions. The structure shown in Fig.~\ref{latticefig} corresponds to two Hamiltonians with an angle of $\gamma$ between the corresponding $\mathbf{n}$ vectors. In order to find an eigenstate for the whole geometry, the solutions have to be fitted at the junctions. We imply Griffith's boundary conditions \cite{G53}, that is, the net spin density current at the junction has to be zero and we also require the continuity of the spinor valued eigenfunctions.

In the case of an infinite periodic structure, we look for Bloch-wave solutions [$\exp(i \mathbf{k\cdot r}) \varphi (\mathbf{r})$ with lattice-periodic spinors $\varphi$)], which means an additional, special boundary condition. The consequence is an energy spectrum with a specific structure: e.g., there will be no solutions in certain energy ranges. Determination of the band structure means finding triples $\{E(\mathbf{k}), k_1, k_2\}$ that correspond to a Bloch-wave eigenspinor of the problem. As usually, we find that $E(\mathbf{k})$ is a multi-valued function of the two dimensional wave vector $\mathbf{k},$ and we can identify infinitely many surfaces in this function. These surfaces (bands) do not overlap unless in the presence of a symmetry induced degeneracy.

Note that the model above assumes single mode propagation, which is a reasonable approximation for narrow conducting wires. Taking the finite width of these channels into account leads to qualitatively the same results, with considerably increased computational costs. Additionally, subbands related to the transversal modes have already been analyzed in detail (see e.g. Ref.~\cite{D95}), thus using the current model we can focus on the band structure induced by the periodicity of the lattice.

\section{III.~Band scheme and conductance properties}
\vspace{-0.5cm}
For a given two-dimensional wave vector $\mathbf{k}$ the energy eigenvalues can be written as $E_{n,m}(\mathbf{k}),$ where the band indices $n$ and $m$ are related to the spatial periodicity of the $\exp(\mathbf{k \cdot r})$ waves in the unit cell along the two lattice directions. These energies scale essentially with the square of $n/a$ and $m/b.$ The same phase relations at the boundaries can hold with e.g. $n$ and $n+1$ waves along the direction of one of the lattice vectors in the unit cell, and the dominant contribution (omitting SOI corrections) of these solutions to the energy is proportional to $n^2/a^2$ and $(n+1)^2/a^2.$ That is, the bands have in general a double quasiperiodic structure. However, when $a\approx b,$ a repetition of a small number of $E_{n,m}(\mathbf{k})$ surfaces provides the complete band structure.

%%%%%%%%%%%%%%%%%%%%%%%%%%%%%%%%%%%%%%%%%%%%%%%%%%%%%%%%%%%%%%%%%%%%%%%%%%%%%%
\begin{figure}[tbh]
\includegraphics[width=8cm]{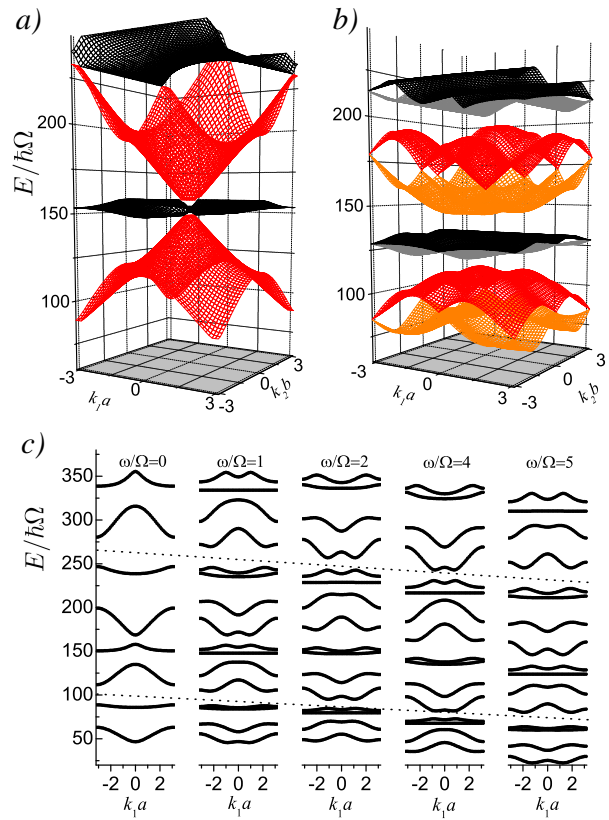}
\caption{Band structure of a lattice with $\gamma=\pi/2$ and $b/a=1.03,$ with strength of the SOI (a) zero and (b)  $\omega/\Omega=5.0.$ Cross sections at $k_2b=1.0$ are shown for several additional values of the SOI strengths in panel c). The thin dotted lines guide the eye by showing the energy range for which the bands  are "essentially the same," i.e., they continuously transform into each other when the SOI strength is changed. Note that the levels between the dotted lines correspond to the two dimensional plots in panels a) and b).}
\label{2dbandfig}
\end{figure}
%%%%%%%%%%%%%%%%%%%%%%%%%%%%%%%%%%%%%%%%%%%%%%%%%%%%%%%%%%%%%%%%%%%%%%%%%%%%%%

Figure \ref{2dbandfig}(a) shows four bands for a rectangular lattice with $b/a=1.03$ and zero SOI.  (Note that energy is measured in units of $\hbar \Omega,$ which, for $a=10$nm in InAlAs/InGaAs based heterostructures is of the order of meV). The four bands seen in this figure are quasiperiodically repeated. The lattices shown in Fig.~\ref{latticefig} have unit cells with four distinct boundary points, that is, four leads connect them to the neighboring cells. The oppositely situated boundary points are equivalent in a crystal, thus any measurable physical quantity has to have the same value at these points. Particularly, the currents carried by the opposite leads should be the same. That is, the sign of the currents at the four leads can be written schematically as $++++,$ $+-+-,$ $-+-+$ and $----$ (where the leads that correspond to the $\pm$ signs follow each other in a clockwise order.) The four bands seen in Fig.~\ref{2dbandfig}(a) correspond to these four possible current configurations.  For nonzero SOI, all these bands split into two due to the spin dependence of the interaction and as it is shown in Fig.~\ref{2dbandfig}, the strength of the SOI modifies considerably both the position and the width of the allowed/forbidden bands.

In Fig.~\ref{2dbandfig}(c) cross sections of the band structure are plotted for different values of the SOI which clearly shows the gradual splitting of the levels as the SOI gets stronger. Additionally, when we identify the bands that continuously evolve from/into each other when the strength of the SOI [characterized by the parameter $\omega$ in Eq.~(\ref{Ham})] is changed, we notice an overall decrease of the energies [see the dotted lines in Fig.~\ref{2dbandfig}(c)]. This is due to the SOI induced splitting of the lowest band, resulting in a decrease of the lowest possible energy when $\omega$ increases.

%%%%%%%%%%%%%%%%%%%%%%%%%%%%%%%%%%%%%%%%%%%%%%%%%%%%%%%%%%%%%%%%%%%%%%%%%%%%%%%%%%%%%%%%%%%%%%%%%%%%%%%%%%%%%%%%%%%%%%%%%%%%%%%%%%%%%%%%%%%%%%%%%%%%%%%%%%%%
\begin{figure}[tbh]
\includegraphics[width=8cm, height=6.3cm]{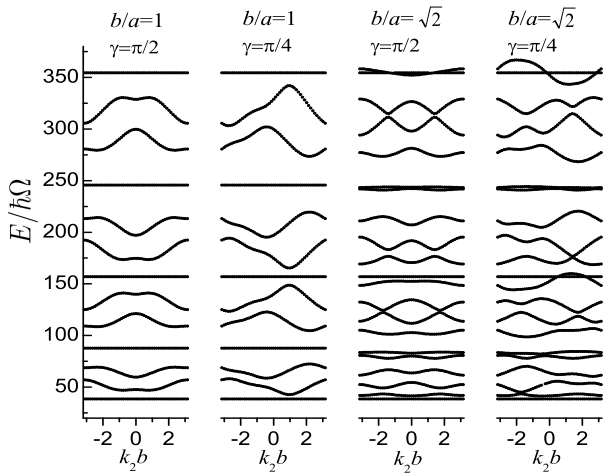}
\caption{The effect of the lattice geometry on the band structure for $\omega/\Omega=1.0,$ $k_1 a=1.0.$}
\label{1dbandfig}
\end{figure}
%%%%%%%%%%%%%%%%%%%%%%%%%%%%%%%%%%%%%%%%%%%%%%%%%%%%%%%%%%%%%%%%%%%%%%%%%%%%%%

The band structure strongly depends on the underlying geometry as illustrated in Fig.~\ref{1dbandfig}. This emphasizes that besides the tunability of the band structure by external gate voltages (that modify the strength of the SOI), geometry is also an important additional degree of freedom.

Results based on infinite structures usually have implications also on large, but finite systems. Now we calculate the conductance of arrays consisting of $N\times N$ unit cells [see Fig.~\ref{latticefig}(b)] using the Landauer-B\"{u}ttiker \cite{D95} formula:
\begin{equation*}
G(E)=\frac{e^{2}}{h} \sum_n \left[T_{\uparrow}^{n}(E) + T_{\downarrow}^{n}(E)\right],
\end{equation*}%
where the sum runs over the possible outputs. $T_{\uparrow}^{n}(E)$ $[ T_{\downarrow}^{n}(E)]$ refers to the transmission probability at the relevant output for spin-up (spin-down) input in the chosen quantization direction. These probabilities are calculated by solving the eigenvalue problem for the whole network at a given energy $E$  imposing the appropriate boundary conditions, e.g.~at the input we have a spin-up (or spin down) incoming wave and a possible reflected one, while at the outputs only outgoing waves appear.

\begin{figure}[tbh]
\includegraphics[width=8cm, , height=11.5cm]{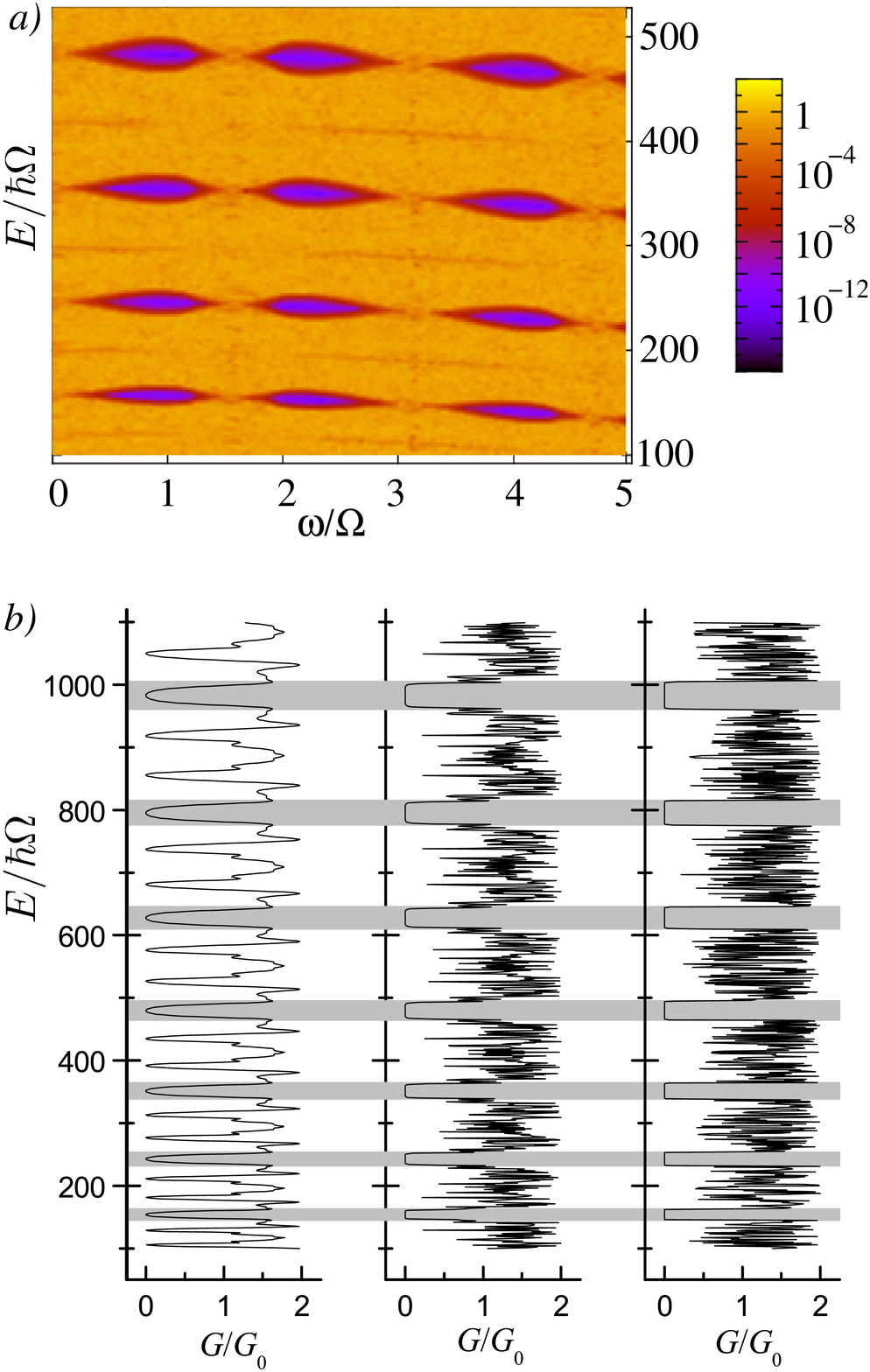}
\caption{Top: Contour plot of the conductance measured in units of $G_0=e^2/h$ for a $15\times15$ array with $b/a=2, \gamma=\pi/2$ as a function of the energy and the SOI strength.  Panel b) shows the conductance of $3\times3$, $7\times7$, $15\times15$ arrays where the light grey shading indicates the energy gaps in the corresponding infinite superlattice.}
\label{finitebandfig}
\end{figure}

Figure \ref{finitebandfig}(a) shows a contour plot of the conductance as a function of the energy and the SOI strength for a rectangular $15\times15$ array. We clearly notice the appearance of stripes (the position and width of which depend on the SOI strength) of zero conductance. In these regions the array is completely opaque for the electrons. Additionally, for the $15\times15$ array, these  "non-conducting stripes" \cite{KFBP08b} coincide with the band gaps obtained from a calculation assuming an infinite structure with the same local geometry. In order to visualize this fact, we projected the band structure on the energy axis to obtain the limits between allowed and forbidden energy regions [see the grey areas in Fig.~\ref{finitebandfig} (b)]. Already for a $3\times3$ network, we can see some signatures in $G(E)$ of the band structure, but for a $7\times7$ array the positions of the zero conductance energy ranges are practically the same as the band gaps. Having introduced random scattering centers as it is discussed in Refs.~\cite{KFBP08b,FKP09}, we observed that the widths of the band gaps decrease only 10\%, even when dephasing is so strong, that 100\% degree of spinpolarization (input) drops to 20\% (outputs).
\vspace{-0.2cm}
%%%%%%%%%%%%%%%%%%%%%%%%%%%%%%%%%%%%%%%%%%%%%%%%%%%%%%%%%%%%%%%%%%%%%%%%%%%%%%
%%%%%%%%%%%%%%%%%%%%%%%%%%%%%%%%%%%%%%%%%%%%%%%%%%%%%%%%%%%%%%%%%%%%%%%%%%%%%%

The most straightforward possible application is energy filtering: E.g., when $a\approx b$ there are very narrow bands
the width and position of which can be controlled by the SOI strength. (For InAlAs/InGaAs based heterostructures and a $9\times9$ network with $a=10$ nm, $b/a=1.005,$ an energy range around 5 meV is transmitted in the middle of a 100 meV wide band gap.)

Additionally, as the width of all the band gaps can be controlled simultaneously by the SOI strength [see the almond-shaped minima in Fig.~\ref{finitebandfig} (a)], an important high temperature effect can be foreseen: even when the input has a broad energy distribution, conductance is modulated by the SOI. For the same device mentioned above, the conductance changes 20\% of its average value when the SOI strength is varied in experimentally achievable range. For a $13\times13$ network the modulation is around 40\%. (Note that in the framework of our model, at "high temperatures" transversal modes other than the ground state in this direction should not be excited. However, the physical reasons of the result above are valid also for multimode propagation.)

For non-square lattices the geometrical anisotropy leads to anisotropy in the band structure (see Fig.~\ref{1dbandfig}), and consequently also in the conductance properties. For a $9\times9$ lattice with $b/a=2, $ $\gamma=\pi/4,$ the difference of the transmission probabilities in the $x$ and $y$ directions -- depending on the SOI strength -- can be zero, or as large as $\pm$0.8, so that the higher one is above 0.95.

Besides the SOI controlled phenomena discussed above, finite arrays can also perform various spin transformations. Apart from spin rotations that can also be done with smaller devices, the arrays considered here are also versatile  spintronic devices: e.g., according to our calculations, the network described in the previous paragraph can deliver oppositely spinpolarized outputs from a completely unpolarized input, when the output leads are situated at the middle of the sides of the network [see Fig.~\ref{latticefig}(b)]. Note that these properties are similar to that of a ring arrays \cite{KFBP08c}, and although in the current case the transmission probabilities are lower than unity (but still around 50\%), the point that there is no need for the local modulation of the SOI strength makes the arrays considered in this paper more promising from the viewpoint of the possible applications.

\section{IV.~Summary and conclusions}
\vspace{-0.5cm}
In this paper we investigated two-dimensional superlattices in which the propagation of the electrons is determined by the interplay of the geometry and the spin-orbit interaction (SOI). We calculated the band structure of these artificial crystals, and showed that by changing the SOI strength in the experimentally achievable range, the band scheme can be modified qualitatively, e.g., forbidden energy ranges can become allowed and vice versa. Comparing the band structure with the conductance properties of finite systems, we found that already for relatively small arrays, forbidden bands are clearly seen in the conductance. This effect -- being based on robust, bulk-like properties -- can be useful for the development of spin-dependent electronic devices. Several possible applications were given, including the strong modulation of the conductance at moderate temperatures.
\smallskip

\textbf{Acknowledgments }
We thank M.~G.~Benedict and F.~Bartha for useful discussions. This work was supported by the
Flemish Science Foundation (FWO-Vl), the Belgian Science Policy (IAP) and the
Hungarian Scientific Research Fund (OTKA) under Contracts Nos.~T81364, M045596. P.F.~was supported by a J.~Bolyai grant of the Hungarian Academy of Sciences.

\end{document}